\documentclass[a4paper,fleqn,usenatbib]{mnras}

\usepackage[T1]{fontenc}
\usepackage{ae,aecompl}

\usepackage{graphicx}	
\usepackage{amsmath}	
\usepackage{amssymb}	

\usepackage{epsfig,graphics,bm}
\usepackage{subfigure}
\usepackage{txfonts}
\usepackage{tabulary} 
\usepackage[utf8]{inputenc}

\title[Dynamic portrait of the Hungaria region]{Dynamic portrait of the region occupied by the Hungaria Asteroids: The influence of Mars}

\author[J. A. Correa-Otto and M. Ca\~nada-Assandri]{
J. A. Correa-Otto\thanks{E-mail: jorgecorreaotto@conicet.gov.ar} and
M.  Ca\~nada-Assandri,\thanks{E-mail: mcanadaassandri@conicet.gov.ar}
\\
Grupo de Ciencias Planetarias, Dpto. de Geof{\'i}sica y Astronom{\'i}a, FCEFyN, UNSJ - CONICET, \\
 Av. J. I. de la Roza 590 oeste, J5402DCS, Rivadavia, San Juan, Argentina\\
}

\date{Accepted XXX. Received YYY; in original form ZZZ}

\pubyear{2018}

\begin{document}
\label{firstpage}
\pagerange{\pageref{firstpage}--\pageref{lastpage}}
\maketitle

\begin{abstract}
The region occupied by the Hungaria asteroids has a high dynamical complexity. In this paper, we analyse the main dynamic structures and their influence on the known asteroids through the construction of maps of initial conditions.  We evolve a set of test particles placed on a perfectly rectangular grid of initial conditions during 3 Myr under the gravitational influence of the Sun and eight planets, from Mercury to Neptune. Moreover, we use the method MEGNO in order to obtain a complete dynamical portrait of the region. A comparison of our maps with the distribution of real objects allows us to detect the main dynamical mechanisms acting in the domain under study such as mean-motion and secular resonances. Our main results is the existence of a small area inside a stable region where are placed the Hungaria asteroids. We found that the influence of Mars has an important role for the dynamic structure of the region, defining the limits for this population of asteroids. Our result is in agreement with previous studies, which have indicated the importance of the eccentricity of Mars for the stability of Hungaria asteroids. However, we found that the secular resonance resulting from the precession of perihelion due to a coupling with that of Jupiter proposed as limit for the Hungaria region could not be determinant for this population of asteroids.
\end{abstract}

\begin{keywords}
celestial mechanics -- methods: numerical -- minor planets, asteroids: dynamical evolution and stability.
\end{keywords}



\section{Introduction}\label{intro}

In the region between Mars and the inner edge of the main asteroid belt there is a dynamical group of asteroids, which is called Hungarias by the first asteroid discovered in the region, (434) Hungaria. The Hungarias are currently clustered in inclinations (16–30$^\circ$) and eccentricities ($e$ < 0.18), but also Williams (1989,1992) and Lemaitre (1994) identified the presence of a family associated to the asteroid (434) Hungaria. Subsequent works (Warner
et al. 2009; Milani et al. 2010) have confirmed a collisional family associated to this asteroid, and they have made an important effort in order to identify which objects of the dynamical group are members of the Hungaria family.

The peculiarity of the location of the Hungaria group has been the reason of numerous dynamical studies seeking to elucidate  the decline rate and the half-life of the population (Migliorini et al. 1998; Milani et al. 2010; McEachern et al. 2010; Cuk 2012; Bottke et al. 2012; Galiazzo \& Schwarz 2014; Cuk \& Nesvorny 2018).. On the other side, observational studies show a population of objects with diameters predominantly greater than $1$ km (McEachern et al. 2010) whose surfaces exhibit a diversity that has been studied both taxonomically and polarimetrically (see e.g., Assandri \& Gil-Hutton 2008; Gil-Hutton et al. 2007; Lucas et al. 2017). Finally, recent studies (e.g., Cuk \& Nesvorny 2018) do not find evidence of a connection between the physical and dynamical properties of the known Hungaria asteroids.

The stability of the Hungaria asteroids population has been studied exhaustively, showing a complex dynamic structure in this region. These studies are especially important to improve our understanding about the origin and the evolution of this population. Previous works (e.g., Milani et al. 2010; Cuk \& Nesvorny 2018) have found an important interaction between Mars and the Hungaria asteroids through close encounters. Moreover, there are several mean-motion resonances (MMRs) and secular resonances (SRs) present in the region. Other results also indicate that the MMRs with Mars and Jupiter, and the SRs with Jupiter and Saturn are the main dynamic modellers of the region. Additionally, in this scenario the non-conservative forces like the dissipative Yarkovsky effect (Farinella \& Vokrouhlicky 1999) are able to affect the temporal evolution of the Hungaria asteroids. The dissipative effects have an influence in the dynamical evolution over a very long time-span, generating a slow chaotic diffusion (McEachern et al. 2010).

All the above-mentioned studies have concentrated on the dynamic evolution of the Hungaria population members, but we do not found in the literature any study about the dynamic portrait of the phase space occupied by this population. Even if the region of the Hungaria asteroids seems to be totally described by the main dynamical structures, a more deeper analysis could reveal some ignored dynamic characteristics that could affect the population, and even provide information about the evolutionary paths of the objects to leave  the region. Thus, the aim of our work is to construct a global dynamic picture of the Hungaria region, combining previous studies (Michel \& Froeschle 1997; Milani et al. 2010; McEachern et al. 2010; Cuk \& Nesvorny 2018) and new results from our investigations. The paper is organized as follows. In Section \ref{map} we describe our dynamical maps, which are building following the current asteroid distribution in the region of interest. A detailed analyse of our results is presented in Section \ref{result}. Conclusions close the paper in Section \ref{conclu}.

\section{Numerical methods: Maps of initial conditions}\label{map}

 \begin{figure}
\raggedleft
 \subfigure{\includegraphics[width=0.43\textwidth]{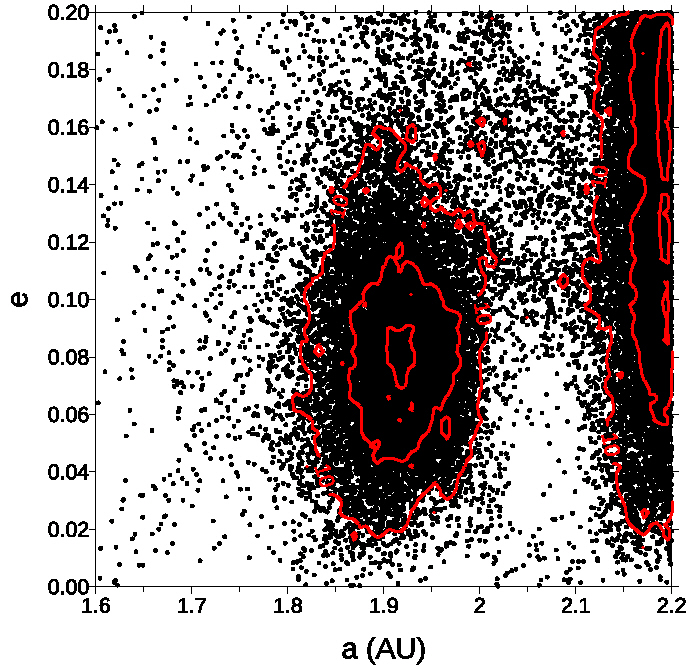}}
 \raggedright
\subfigure{\includegraphics[width=0.47\textwidth]{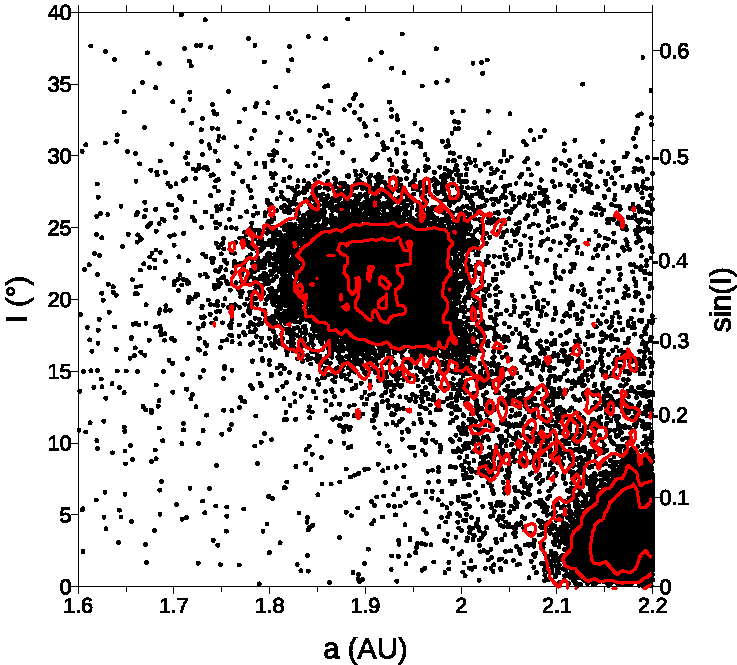}}
\caption{Osculating orbital elements: semimajor axis, eccentricity and inclinations for the $\sim 30000$ asteroids currently known  close to the Hungaria region (ftp://ftp.lowell.edu/pub/elgb/astorb.html), updated at October 2017. In this plot we show their distribution in the $a$-$e$ plane (top panel) and the $a$-$I$ ($a$-$\sin{I}$) plane (bottom panel), in the limits: 1.6 $<$ $a$ $<$ 2.2 AU, $e$ $<$ 0.2 and $I$ $<$ 40$^\circ$. In red line, we have plotted level curves of density. For each plane we plot three representative curves, for the $a$-$e$ plane  we plot that corresponding to region with more than 10, 40 and 80 objects by bin, and for the $a$-$I$ plane the levels correspond to values 1, 5 and 16.}
\label{fig0a}
\end{figure}

The spatial distribution of the asteroids in the Hungaria region makes necessary a  dynamical study in more than one plane, being the most usual the $a$-$e$ and the $a$-$I$ ($a$-$\sin{I}$) planes. The accepted natural borders of the Hungaria group is  between $1.78$ and $2.06$ au, $e<0.18$ and inclinations between $16$ and $30$ degrees. These dynamical edges are bounded by the $\nu_5$ and $\nu_{16}$ SRs, the 4:1 MMR with Jupiter and the orbit of Mars (Gradie et al. 1979; Milani et al. 2010). Although, the 2:3 MMR with Mars has recently been proposed as an upper limit for the semimajor axis  (Warner et al. 2009).

Then, we analyse an extended area defined by the osculating semimajor axis range $a\in (1.6, 2.2)$ au, and the limits $e<0.2$ and $I< 40^\circ$, which contain about $3 \times 10^4$ asteroids. The distribution of the osculating orbital elements of the asteroids in this region is shown in Figure \ref{fig0a}. However, a better way to visualize their distribution was proposed by Michtchenko et al. (2010) who suggested to use the density distribution of the asteroids in each plane, which is show as red level curves in Figure \ref{fig0a}. The density distribution  was calculated dividing both planes in a rectangular grid of $50$ by $50$ cells, and calculating the number of real objects inside each cell to plot the level curves of density. In the planes we plot only three lines of density for simplicity: in the $a$-$e$ plane we represent the lines of density corresponding to $10$, $40$ and $80$, while in the $a$-$I$ plane we plot the lines corresponding to $1$, $5$ and $16$. In each plane, the region outside the first limit represent the space of low density, the intermediate regions have a density value between the inner and outer limits, and the central zone is the more dense region. It is possible to see that the spatial density distribution shows a concentration of objects in a region with mean values: $\sim 1.9$ au, $\sim 0.8$  and $\sim 20^\circ$  for $a$, $e$ and $I$, respectively. Moreover, the concentration of objects shows a density bulge in the $a$-$e$-$I$ space in a box with approximate limits $a\in (1.85,2.00)$ au, $e\in (0.04,0.12)$ and $I\in (16^\circ,25^\circ)$.

We study the dynamic structure of the phase space occupied by the Hungaria asteroids using fictitious test particles and constructing stability maps of initial conditions, where each test particle is placed over a perfectly rectangular grid.

Therefore, to improve our understanding of the main dynamic mechanisms acting in this region we study the dynamic structure of the phase space occupied by the Hungaria group using fictitious test particles and stability maps. We consider only gravitational interactions because the integration time-span of our maps allow us to ignore in a first approximation the effect of non-conservative forces. We have constructed two pairs of maps with planes $a$-$e$ and $a$-$I$ ($a$-$\sin{I}$) in the range $1.6 - 2.2$ au for the osculating semimajor axis, $e$ $\in (0.0,0.2)$ and $I$ $\in (0^\circ,40^\circ)$. In the first pair of maps we placed $1.2 \times  10^6$ massless particles with separations $\Delta a = 10^{-3}$ au and $\Delta e = 10^{-3}$/$\Delta I = 0.2^\circ$, which were integrated for $T = 0.1$ Myr  and analysed with MEGNO (Mean Exponential Growth Factor of Nearby Orbits, Cincotta \& Simo 2000). The second pair of maps were constructed using a grid of $100$ by $40$ cells with separations of  $\Delta a = 6 \times 10^{-3}$ au in semimajor axis and $\Delta e = 5 \times 10^{-3}$/$\Delta I = 1^\circ$ in eccentricity/inclination, and in this case we analyse the survival times after $T = 3$ Myr.

For all the maps the initial osculating angles were chosen to be equal to zero. Although the arbitrary choice of angles can make us lose the possibility of detecting certain specific dynamic properties, the main dynamical structures still are present in our maps. Finally, we define the initial value of inclination/eccentricity for the maps in the $a$-$e$/$a$-$I$ plane from the distribution of Hungaria asteroids. The bulk of real objects of Fig. \ref{fig0a} has as most probable values $e \sim$ 0.08 and $I \sim$ 20$^\circ$, which are chosen as initial values for the $a$-$I$ plane and the $a$-$e$ plane, respectively

The temporal evolution of each initial condition in the four maps were solved through numerical integrations of the exact equations of motion with the \textit{N-corp} integrator (Correa Otto et al. 2010), which employs the Bulirsch-Stoer code with adopted accuracy of $10^{-13}$, and it has the option to calculate MEGNO and/or survival time-span for each massless particle. In the simulation we include the gravitational interactions of the Sun and the eight planets, from Mercury to Neptune, and we obtain their orbital elements from the JPL database of the NASA (http://ssd.jpl.nasa.gov).

 \begin{figure}
 \centering
\includegraphics[width=0.45\textwidth]{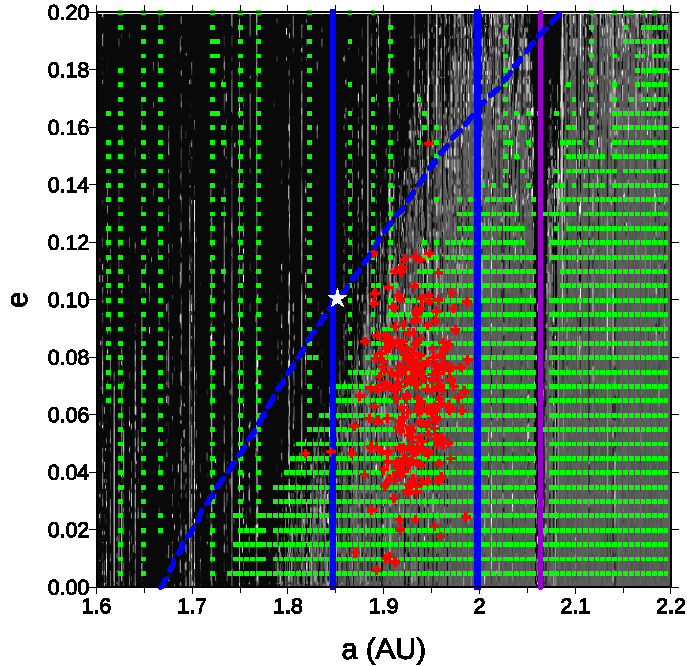}
 \caption{{\small Dynamic maps in the plane of osculating semimajor axis and eccentricity.  The grey-scale levels correspond to the MEGNO-map,  white colors indicate periodic motion, lighter regions correspond to regular motion and darker tones indicate increasingly chaotic motion. The survival-map is overlap, where the green dots  indicate initial conditions that survive 3 Myr. For details about the maps, see the text. The purple continuous line indicate the 4:1 MMR with Jupiter, and the blue continuous lines show the 3:4 and 2:3 MMR with Mars (1.845 and 1.997 AU respectively). Blue dashed line indicate the aphelion distance of Mars. Red crosses correspond to the position of $\sim$ 500 Hungaria asteroids in the maps, and the white star shows the position of the asteroid (156466) \textit{2002 CG10}, which is captured in the 3:4 MMRs with Mars.}}
\label{map1}
\end{figure}

Finally, in order to obtain a more complete picture of the problem we compare the main dynamic structures in the phase space with the position of the real objects. Thus, we follow the temporal evolution of $\sim 5000$ Hungaria asteroids for $1$ Myr searching for the objects which survive all the integration time and cross some of the two planes. We save the position of the objects when the angles are close to zero with a tolerance of 0.5$^\circ$, and the inclination/eccentricity is close to 20$^\circ$/0.08 with a tolerance of 0.5$^\circ$/0.005 for the $a$-$e$/$a$-$I$ plane. We have found a total of $\sim 800$ objects crossing one or both planes, which represent approximately a $15\%$ of the total sample. There are about $500$ objects crossing each plane, with $\sim 200$ asteroids that cross both planes.

\section{Results}\label{result}

The aim of the first pair of maps, MEGNO-maps hereafter, is to give us a detailed dynamical analysis of the region. The large number of MMRs in the interval of semimajor axis considered in the maps (Milani et al. 2010; McEachern et al. 2010) made necessary a grid of high precision.  The values of the averaged MEGNO number is $0$ for periodic orbits, regular orbits yield to values $\leq 2$, while larger values are indicative of chaotic motion. In the maps, the calculated values of MEGNO, in the range from $0$ to $2.1$ and higher, are coded in grey-scale varying from white (0) to black ($\geq 2.1$).

In the other two maps, survival-maps hereafter, we lost resolution but we won in integration time-span, which is important because our aim is to confirm the prediction of stability of the MEGNO-maps. Then, for the survival-maps we indicate the particles that survive the complete integrated time-span of $3$ Myr by a green dot, and we superpose them to the MEGNO-maps.

Our results are shown in Figures \ref{map1} and \ref{map2}, where we can see that with the exception of particular areas in the maps (high eccentricity in the $a$-$e$ plane, and low inclination for small $a$ in $a$-$I$ plane), the predictions of MEGNO-maps are in agreement with the results obtained in the survival-maps. Moreover, we include the position of the $\sim 800$  real asteroids ($500$ in each map) with red/white crosses in the $a$-$e$/$a$-$I$ plane.

The main dynamic feature find in the maps is a region dominated by chaos (Fig. \ref{map1} and \ref{map2}). There is a stable area located  at the right border of the map, which correspond to the inner main belt. This area is interrupted by the 4:1 MMR with Jupiter at $a\sim 2.06$ au, which leave a reduced portion of the stable area in the center of the map where the  Hungaria asteroids can survive for long periods.

\begin{table}
\caption{Nominal semimajor axis of several MMRs with Jupiter, Saturn and Earth, in the Hungaria region. Semi-major axes are rounded to the 3rd decimal, and the two last columns indicate order and grade.}
\label{table1}
\centering
\begin{tabular}{c c c c c  }
\hline\hline
  Planet    &  MMR   & $a$ (au) & order  & grade  \\
\hline
Jupiter &	17:3 &	1.636 &	14 & 3  \\
Saturn &	14:1 &	1.641 &	13 & 1  \\
Jupiter &	11:2 &	1.669 &	9 & 2  \\
Saturn &	27:2 &	1.682 &	25 & 2  \\
Jupiter &	16:3 &	1.704 &	13 & 3  \\
Saturn &	13:1 &	1.725 &	12 & 1  \\
Saturn &	25:2 &	1.770 &	23 & 2  \\
Jupiter &	5:1 &	1.779 &	4 & 1  \\
Earth &	5:12 &	1.793 &	7 & 5   \\
Saturn &	12:1 &	1.819 &	11 & 1  \\
Earth &	2:5 &	1.846 &	3 & 2   \\
Jupiter &	14:3 &	1.863 &	11 & 3  \\
Saturn &	23:2 &	1.872 &	21 & 2  \\
Jupiter &	9:2 &	1.908 &	7 & 2  \\
Earth &	3:8 &	1.923 &	5 & 3   \\
Saturn &	11:1 &	1.928 &	10 & 1  \\
Jupiter &	13:3 &	1.958 &	10 & 3  \\
Earth &	4:11 &	1.963 &	7 & 4   \\
Saturn &	21:2 &	1.989 &	19 & 2  \\
Saturn &	10:1 &	2.054 &	 9 & 1  \\
Jupiter &	4:1 &	2.064 &	3 & 1  \\
Earth &	1:3 &	2.080 &	2 & 1   \\
Jupiter &	11:3 &	2.187 &	11 & 3  \\
Saturn &	19:2 &	2.126 &	17 & 2  \\
\hline
\end{tabular}
\end{table}

\subsection{The $a-e$ plane: The influence of the mean-motion resonances and the Martian eccentricity}\label{mmr}

The main relevant feature in the maps (Fig. \ref{map1}) is the occurrence of several vertical stripes of stable/chaotic motion, which are associated to MMRs with the planets from Earth to Saturn. This structure of vertical stripes is in agreement with the results of previous works (Milani et al. 2010; McEachern et al. 2010). In Tables  \ref{table1} and  \ref{table2} we indicate the most important MMRs with Earth, Jupiter, Saturn and Mars. The high density of MMRs in the region form a forest of lines (e.g., Leiva et al. 2013), which  make impossible to indicate all of them in the map, but due to their importance to our analysis we only include the position of the 4:1 MMR with Jupiter ($\sim 2.06$ au) in purple line, and the 3:4 and 2:3 MMR with Mars ($\sim 1.85$ and $2.00$ au, respectively) in blue lines.

\begin{table}
\caption{Same as Table \ref{table1}, but now for the MMRs with Mars.} 
\label{table2}
\centering
\begin{tabular}{c c c c c  }
\hline\hline
  Planet    &  MMR   & $a$ (au) & order  & grade  \\
\hline
Mars &	6:7 &	1.688 &	1 & 6  \\
Mars &	5:6 &	1.721 &	1 & 5  \\
Mars &	17:21 &	1.754 &	4 & 17  \\
Mars &	4:5 &	1.768 &	1 & 4  \\
Mars &	19:24 &	1.780 &	5 & 19  \\
Mars &	15:19 &	1.784 &	4 & 15  \\
Mars &	11:14 &	1.787 &	3 & 11  \\
Mars &	18:23 &	1.793 &	5 & 18  \\
Mars &	7:9 &	1.802 &	2 & 7  \\
Mars &	17:22 &	1.809 &	5 & 17  \\
Mars &	10:13 &	1.815 &	3 & 10  \\
Mars &	13:17 &	1.822 &	4 & 13  \\
Mars &	16:21 &	1.826 &	5 & 16  \\
Mars &	3:4 &	1.846 &	1 & 3  \\
Mars &	14:19 &	1.867 &	5 & 14  \\
Mars &	11:15 &	1.874 &	4 & 11  \\
Mars &	8:11 &	1.884 &	3 & 8  \\
Mars &	13:18 &	1.893 &	5 & 13  \\
Mars &	5:7 &	1.907 &	2 & 5  \\
Mars &	12:17 &	1.922 &	5 & 12  \\
Mars &	7:10 &	1.933 &	3 & 7  \\
Mars &	9:13 &	1.947 &	4 & 9  \\
Mars &	11:16 &	1.956 &	5 & 11  \\
Mars &	2:3 &	1.997 &	1 & 2  \\
Mars &	9:14 &	2.045 &	5 & 9  \\
Mars &	7:11 &	2.060 &	4 & 7  \\
Mars &	5:8 &	2.084 &	3 & 5  \\
Mars &	8:13 &	2.106 &	5 & 8  \\
Mars &	3:5 &	2.141 &	2 & 3  \\
\hline
\end{tabular}
\end{table}

The importance of the MMRs for the dynamics of the region can be appreciated in the survival-map, where the vertical stripes are still present in the chaotic area ($a\leq 1.85$ au) despite the low resolution of the map. This is because the MMRs are able to modify the stability of the region, yielding a local island of stability in a chaotic sea or, conversely, strips of unstable motion in stable areas (e.g., 4:1 MMR with Jupiter at $2.06$ au).

\begin{figure}
 \centering
\includegraphics[width=0.45\textwidth]{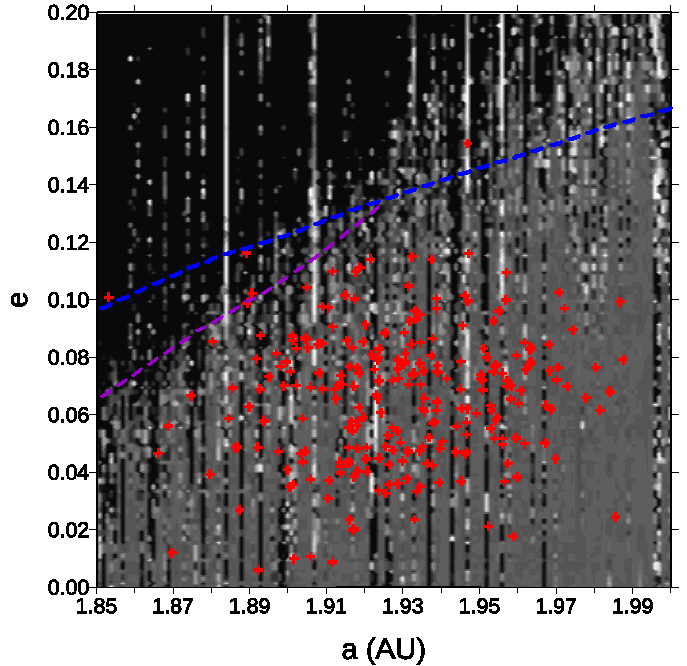}
\caption{Same as Figure \ref{map1}, but now zooming the MEGNO-map in the region between the 3:4 and 2:3 MMRs with Mars. Notice the increase of the width and the overlap of the resonances with $e$. Blue and purple dashed lines show the aphelion distance of Mars and the stability limit by overlap of adjacent MMRs, respectively. }
\label{figmars}
\end{figure}

Moreover, at high eccentricities the forest-like structure of resonances affect the region through the  overlap of MMRs (e.g., Chirikov 1979; Wisdom 1980; Ramos et al. 2015). The width of MMRs depends on the eccentricity (Murray \& Dermott 1999), which we can see in the 4:1 MMR with Jupiter. Fig. \ref{figmars} shows an enlargement of the region between the 3:4 and 2:3 MMRs with Mars, where we can see the same behaviour for the width of the several resonances in the area. Then, for adjacent or close MMRs the increase of their widths result in an overlap at high values of $e$. The criterion of Wisdom (1980) for an overlap is when a separatrix of one resonance has crossed a separatrix of the other resonance, such that the combined local chaos of each separatrix become in a global chaos able to produce orbital instability. Therefore, due to the overlap of MMRs we can see the vertical structure of the stability region with a saw-tooth shape (see, Ramos et al. 2015)  for $a < 2$ au.

Another important issue about the MMRs is to determine  when an asteroid is effectively captured in a resonance. An object placed close to the nominal value of a MMR does not ensure its capture inside the resonance. Even more, the temporal evolution of the semimajor axis is not a proof of evolution inside the MMR. For example, for low values of eccentricity there  are cases of quasi-resonant regime of motion  where the objects are not in a resonance (see, Correa-Otto et al. 2013). The only way to confirm the evolution of the asteroids inside of such narrow resonances is by an analysis of the resonant angle.

\begin{figure}
 \centering
\includegraphics[width=0.45\textwidth]{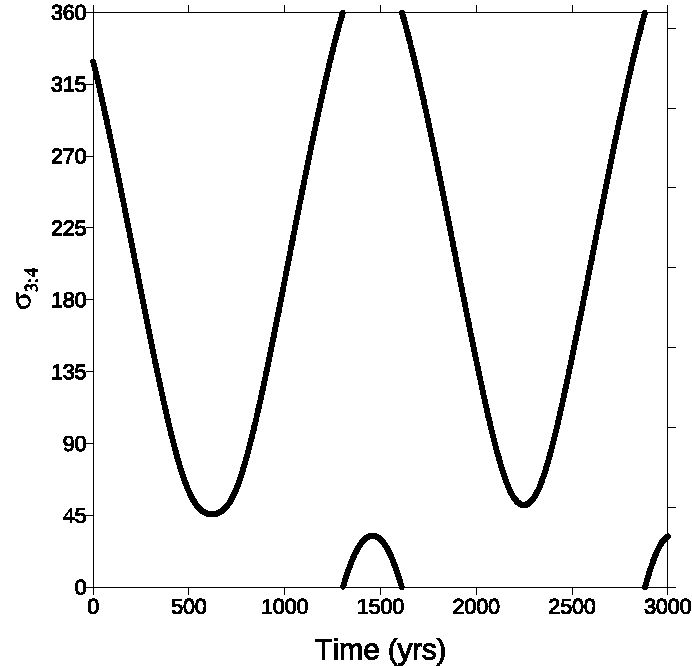}
 \caption{Temporal evolution of the resonant angle of the Hungaria asteroid (156466) \textit{2002 CG10}, which is captured in the 3:4 MMR with Mars (librating around an asymmetric solution).}
\label{fig8}
\end{figure}

The asteroid (156466) \textit{2002 CG10} can be considered as an example of the importance of the analysis of the resonant angle. The position of this asteroid in the $a$-$e$ plane is indicated with a white star in Fig. \ref{map1}, where we can see that to study the resonant dynamic of this object is not enough an analysis of the semimajor axis because the 2:5 MMR with Earth and the 3:4 MMR with Mars overlaps at the same nominal position (see Tables \ref{table1} and \ref{table2}). The temporal evolution of both resonant angles show a circulation corresponding to the MMR with the Earth, and an asymmetric libration (see, Michtchenko et al. 2006) of the Martian resonant angle, which is shows in Figure \ref{fig8}. The dynamic evolution of this asteroid is complex because it is in the limit of the stable motion (even more, it is a Mars crossing), but the resonance protects it from the chaos, allowing to survive by $\sim 1$ Myr before being destabilized by Mars. It is worth to note that although there are 17 asteroids in the 2:3 MMR (Connors et al. 2008), this asteroid is the first reported object captured in the 3:4 MMR with Mars.

On the other hand, we can see that there are boundaries for the Hungaria region. From our results (Fig.  \ref{map1}) we found an upper limit in $a$ and $e$ for the stable zone: the $a$-limit correspond to the 4:1 MMR with Jupiter and the $e$-limit is associated to the initial conditions that their orbits are crossing that of Mars. (Milani et al. 2010; Cuk \& Nesvorny 2018). The blue dashed line in Fig.  \ref{map1} and \ref{figmars} indicates where the perihelion distance of the particles are at the same aphelion distance of Mars. 

However, the population of Hungaria asteroids is not distributed along the stable zone, in fact the real objects (red crosses) occupy a reduced region defined by other dynamical boundaries. The sample of asteroids that cross the $a$-$e$ plane shows a bulk of objects contained between the 3:4 and 2:3  Martian MMRs, with a sharp drop in density beyond this resonances. Moreover, the limit in eccentricity seems to be different to that defined by the direct interaction with Mars, for $a < 1.92$ au the stability limit in $e$ departs from the smooth curve of aphelion, and we can see that the saw-tooth structure defining that limit. In Fig. \ref{figmars} is possible to appreciate the effects of the MMRs overlap which modify the stability region and define a new $e$-limit, which is qualitatively indicated with a purple dashed line in that figure. The region has a lot of MMRs with several planets and to calculate the overlap of all these resonances and the exact position of the limit is a very complex task, which we let for a future work.

Then, the comparison of our maps with the asteroids (red crosses) suggest that the lower and upper limits in $a$ for the Hungaria asteroids are the 3:4 and 2:3 MMRs with Mars, respectively, while the 4:1 MMR with Jupiter defines the limit of the stable area. The gap in the population of asteroids in the stable area between the 2:3 MMR with Mars and the 4:1 MMR with Jupiter could have important consequences for the hypotheses about the migration of objects from the main belt to the Hungaria region.

Additionally, our results complement that of Cuk \& Nesvorny (2018), because while for $a>$ 1.95 au the limit in eccentricity seems to be defined by the direct interaction between the Hungaria objects and Mars, for the region closest to Mars (i.e., $a<$ 1.95 au) the overlap of MMRs results a better indicator of the $e$-limit. It is worth to note that, the manifestation of the orbital instability by the overlap of MMRs take place in less than $0.1$ Myr, which represent a fast destabilizing phenomenon similar to the direct interaction with Mars. Even so, our result is in agreement with that of Cuk \& Nesvorny (2018) about the influence of the Martian eccentricity ($e_{mars}$) on the $e$-limit, because the $e_{mars}$ affects the width of the Martian MMRs and hence the overlap of resonances.

Finally, it is possible to deduce that for a longer time of integration if non-conservative forces, like the Yarkovsky effect, are considered the objects will be able to  enter or cross MMRs. When an asteroid is inside a MMR or has a fast passage through a resonance, its eccentricity is excited and increased (Roig et al. 2008; Brown et al. 2007; Gallardo et al. 2011). This dynamic evolution in the Hungaria region could lead the objects to cross the Martian orbit, or to reach the chaotic area of overlap of MMRs.  Then, it is possible to predict a slow decrease of the population because of this mechanism.

\subsection{The $a-I$ plane: The influence of the secular resonances}

 \begin{figure}
 \centering
\includegraphics[width=0.45\textwidth]{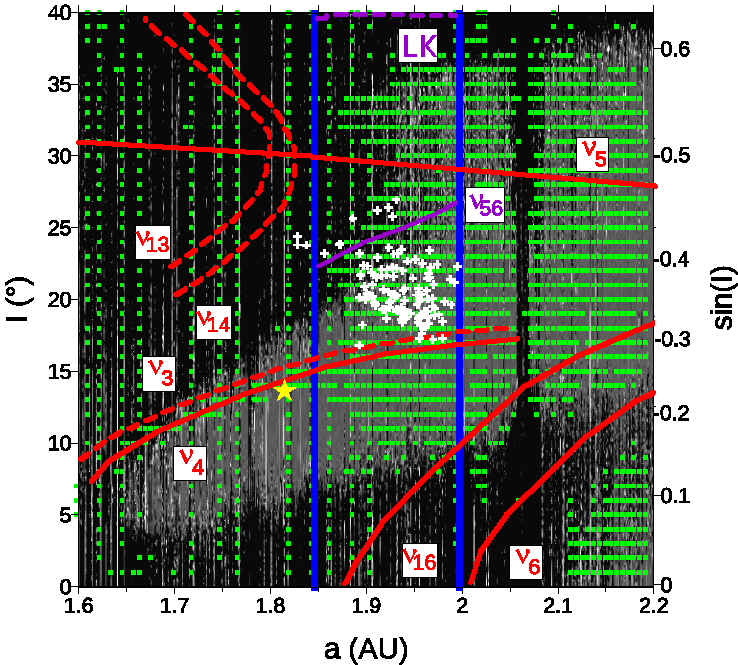}
 \caption{{\small Same maps as in Fig. \ref{map1}, except that for the $a$-$I$ ($\sin{I}$) plane. The blue continuous lines indicate the 3:4 and 2:3 MMR with Mars. Linear secular resonances $\nu_i$ and $\nu_{1i}$ from Earth to Saturn (i.e., $i=$ 3 to 6) are in red line, and the nonlinear $\nu_{56}$ SR is in purple line. Moreover, we include 500 Hungaria asteroids that cross the map in white crosses, and in yellow star we show the position of the asteroid (244666) \textit{2003 JT1}, which is in the $\nu_4$ SR with Mars.}} 
\label{map2}
\end{figure}

Figure \ref{map2} shows the $a$-$I$ plane, where the 2:3 and 4:3 MMRs with Mars are indicated with blue vertical lines. We can see that in addition to the vertical stripes associated to MMRs, it is possible to appreciate other dynamic structures modulating the stability of the region, which correspond to secular resonances.

There are many information in the literature about the position of the main secular resonances in the Hungaria region (Williams \& Faulkner 1981; Michel \& Froeschle 1997; Milani et al. 2010). However, a problem for our analysis is that the SRs are represented in the plane of proper orbital elements (Morbidelli \& Henrard 1991; Knezevic et al. 1991; Froeschle \& Morbidelli 1994), and it is very difficult to represent them in our maps. In the plane of osculating elements the SRs is not a single line, as in the case of proper elements, and occupy a region which is also affected by the presence of unstable zones.

Our method to identify the area of influence of SRs consists in following the temporal evolution of the resonant angles searching for the cases where a libration appears for the initial conditions in the survival-map. So, we considered:
\begin{itemize}
\item Four linear $g$-type  SRs (i.e. precession of perihelion) corresponding to the planets Earth ($\nu_3$), Mars ($\nu_4$), Jupiter ($\nu_5$) and Saturn ($\nu_6$).
\item Three linear $s$-type  SRs (i.e. precession of node) corresponding to the planets Earth ($\nu_{13}$), Mars ($\nu_{14}$) and Saturn ($\nu_{16}$).
\item  The nonlinear SR $\nu_{56} = g - s - g_5 + s_6$ with Jupiter and Saturn, where $g$, $s$, $g_5$ and $s_6$ are the osculating frequency of perihelion and the osculating frequency of the longitude of the node of the asteroid and the planets.
\item The Lidov-Kozai resonance (LK resonance hereafter, Lidov 1961; Kozai 1962) with Mars
\end{itemize}
Then, in Fig. \ref{map2} we schematically draw the approximated position of the $g$-type and $s$-type resonances with red lines, and the nonlinear $\nu_{56}$ SR and the LK resonance are indicated by a continuous and dashed purple line, respectively.

The linear SRs with Earth and Mars are at medium and high values of inclination. The $\nu_{13}$ and $\nu_{14}$ SRs are in the chaotic region, where we can not follow the temporal evolution of the resonant angles. So, from previous results on the proper orbital elements (Michel \& Froeschle 1997; Milani et al. 2010) we qualitatively draw such SRs with dashed lines. The $\nu_4$ SR with Mars appears like a stable appendix structure between 8$^\circ$ and 15$^\circ$, below $\sim 1.85$ au, and it extends through the stable region until the 4:1 MMR with Jupiter. This result shows the  strength of the  $\nu_4$ SR, which allows the survive of initial conditions very close to the orbit of Mars through the stable appendix. However, the dynamical effects of the  $\nu_3$ SR seems to be masked by the influence of Mars, because it should appear close to the $\nu_4$ resonance but we do not find initial conditions with libration of such angle. From previous studies (Michel \& Froeschle 1997) we  indicate its position with a dashed red line close to the $\nu_4$ SR in Fig. \ref{map2}. This dynamic situation is similar to that produced by the 2:5 and 2:3 MMRs with the Earth and Mars described in a previous section, and allows to conclude that in the studied area the influence of Mars is stronger than the effects of the Earth. Moreover, similar to the 3:4 MMR, we found an object evolving in the $\nu_4$ SR: the asteroid (244666) \textit{2003 JT1}, which is indicated by a yellow star in Figure \ref{map2}, which is reported for the first time and it is the only know case of an asteroid captured in a SR with Mars in the Hungaria region.  The Figure \ref{fig4} shows the temporal evolution of the resonant angle for this object, were we can see the period of $\sim 18000$ years for the secular resonance. The object is not close to any MMRs and  it is a Mars crossing which survives the numerical integration during $0.5$ Myrs and has $5$ close encounters with Mars before this planet destabilized it.

\begin{figure}
 \centering
\includegraphics[width=0.45\textwidth]{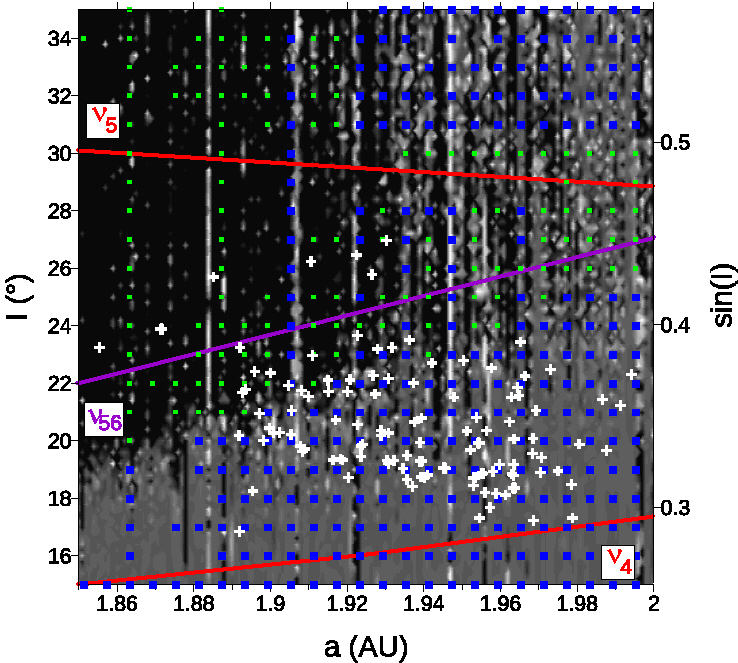}
 \caption{{\small Same as Fig. \ref{map2}, but now zooming the region between the 3:4 and 2:3 martian MMRs and $I$ $\in$ (15,35)$^\circ$. The blue dots indicate initial conditions that survive 50 Myr. }} 
\label{map3}
\end{figure}

\begin{figure}
 \centering
\includegraphics[width=0.45\textwidth]{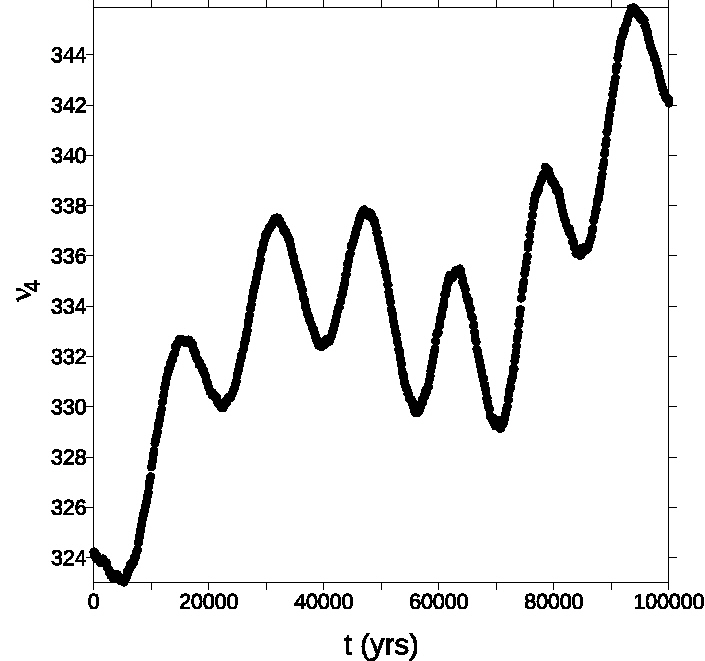}
\caption{Temporal evolution of the resonant angle of the Hungaria asteroid (244666) \textit{2003 JT1}, which is captured in the $\nu_4$ SR with Mars  (librating around an asymmetric solution).}
\label{fig4}
\end{figure}

The $\nu_{6}$ and  $\nu_{16}$ SRs with Saturn appear at low inclinations and the $\nu_5$ SR with Jupiter is found at $\sim$ 30$^\circ$, their areas of influence are in agreement with the result of previous works (Michel \& Froeschle 1997; Milani et al. 2010). The SRs with Saturn are responsible by the chaotic regions at low inclinations observed in our map, and the disturbing effects of the SR with Jupiter appear in a small area in the survive-map. Moreover, the nonlinear $\nu_{56}$ SR is found at $\sim$ 23$^\circ$, although its dynamical influence is not observed in the survival-map because the short time span of our simulations. Milani et al. (2010) have found a period of $\sim$ 3.5 Myr for the  $\nu_{56}$ SR, so in order to appreciate its influence and confirm the MEGNO-map predictions, we increase the integration time in the area of the survive-map corresponding to that resonance. For the particles in the box defined by the 3:4 and 2:3 MMR  with Mars and I $\in$ (15,35)$^\circ$ we integrate by 50 Myr, which corresponds to a $\sim$ 15 periods of the $\nu_{56}$ SR. The results are in Figure \ref{map3}, where we can see that now the survival initial conditions are in agreement with the MEGNO-map and it is possible to appreciate the influence of the nonlinear resonance.

We do not find in the literature studies about the chaotic area with $I> 35^\circ$ in the Hungaria region, and the high values of the inclination make the LK resonance one of the main secular perturbations to  be considered. Although the previous result of Michel \& Froeschle (1997) suggest that the influence of the LK resonance is out of the scope of our maps, there is a  complex dynamic structure in the region $a \in$ (1.85, 2.00) au and we find that the temporal evolution of the argument of perihelion ($\omega$) of most of the test particles with $I \geq 39^\circ$ alternates between circulation and libration, which confirm the influence of the LK resonance with Mars in such region.  

In reference to the boundaries of the Hungaria region, the $a$-limits were discussed in previous section, so we only analyse here the limits in inclination (i.e. $I$-limits). The $\nu_6$ and $\nu_{16}$ SRs play the role of bottom $I$-limit for the stable area, while the upper stability $I$-limit is defined by the secular perturbation of the LK resonance. The orbital instability at such high values of Inclination could be the result of the combination of the dynamic effects in the area: the secular perturbation of the LK resonance excite the eccentricities of the particles to high values, where the Martian aphelion distance and the overlap of MMRs become important. This chaotic area is very interesting to do a detailed study which we are already doing, and whose results will be left for a future paper because it is out of the scope of the present study.

However, we found a similar scenario to that of the $a$-$e$ plane, where the Hungaria asteroids are not all distributed along the stable area, instead the bulk of density in the population has well defined boundaries. The $\nu_4$ SR modulates the lower limit in $I$ for the real objects (white crosses) at $I \sim 15^\circ$, and once again the influence of Mars results very important for the population of the Hungaria asteroids. On the other hand, the upper $I$-limit is associated to the nonlinear $\nu_{56}$ SR, where we can see a sharp drop in the density of real objects (white crosses). It is worth to note that we find very few objects between the nonlinear resonance and  $\nu_{5}$ SR, which indicate a small influence of this linear resonance with Jupiter for the population and for the region. This result is opposite to that of previous works, where the SR with Jupiter is considered as an important dynamic modulator of the Hungaria population.

Finally, when we compare our results in the plane of osculating elements, with the distribution in the plane of synthetic proper elements presented in Fig. 8 of Milani et al. (2010), we confirm our results for the $I$-limits. In the plane of synthetic proper elements the $\nu_4$ SR also seems to be the lower limit, and there are a low density of object over the $\nu_{56}$ SR.  Then, it is worth to note that if the region of the Hungarias is replenish with objects coming from the main asteroid belt, the flux of material has to income with an intermediate inclination, because there is a small probability to find objects at low or high values of inclination. This represent a challenge for the comprehension of the origin of the  Hungaria asteroids and their evolution.

\section{Conclusions}\label{conclu}

In this paper we have presented a dynamical portrait of the extended Hungaria region. The aim of our work was to identify the main dynamic structures that modulate the density distribution of the Hungaria asteroids in order to improve our understanding about the complexity of this region. However, we have present a different approach from previous work studying the phase space of the population through the analyse of a grid of initial conditions of fictitious particles instead of following the evolution of the asteroids.

Our main result is the determination of the existence of two regions in the phase space occupied by the Hungarias asteroids. There is a stable area defined by the classical limits and there is a small region contained in this stable area where we found the population of asteroids. This small region has well defined limits, which have Mars as main modulator. It is worth to note that, even though our results is obtained in the plane of osculating elements, the distribution in the plane of synthetic proper elements  shows similar boundaries Milani et al. (2010).

Although some of the limits for the stable area are in agreement with previous works, we found other dynamic structures modulating the region. The classical limits of the 4:1 MMR with Jupiter in $a$, the Mars crossing in $e$, and the $\nu_6$ and $\nu_{16}$ SRs with Saturn as lower limit in $I$ are in agreement with previous results. However, we also found an important influence of the overlap of MMRs for the $e$-limit, and the LK SR with Mars acts like an upper limit in $I$. On the other hand, the effective limits for the population of Hungaria asteroids in the $a$-$e$-$I$ space are the 2:3 and 3:4 MMRs with Mars in semimajor axis, the direct interaction with Mars plus the overlap of MMRs in eccentricity, and nonlinear $\nu_{56}$ SR and the $\nu_4$ SR with Mars in inclination. Some of these limits have already been indicated, so our result is in agreement with previous works.

From the obtained results, we can deduce three important consequences in the dynamic study of the Hungaria asteroids. First, we found that the phenomena of overlap of MMRs have an important influence in the orbital stability of the objects in the region. This result complements that of Cuk \& Nesvorny (2018), because the instability by direct interaction with Mars (i.e. Mars crossing) is increased by the phenomena of overlap of resonances. Moreover, the Martian eccentricity has an important role for the $e$-limit due to its influence in the aphelion distance and in the phenomena of overlap, which is in agreement with the results of Cuk \& Nesvorny (2018).

Second, we found that the $\nu_5$ SR appears as a perturbation in the stable area and it is not an upper limit in $I$, which is associated to the Martian LK resonance. Even more, this resonance with Jupiter does not seem to be important for the Hungaria population. This result is opposed to previous works where this resonance has been indicated as very important for the Hungarias asteroids. So, we can conclude that in such dynamicaly complex region it is necessary to analyse the phase space plus the population in order to obtain a more complete appreciation of the main dynamical structures and their importance.

The third and last point is the importance of our study for the evolution of the objects in the region. Our results suggest a mechanism to empty the Hungaria region since through the combined effects of MMRs, SRs and non-conservative forces (i.e., Yarkovsky effect), the asteroids are able to increase their eccentricities or inclinations until their orbits become unstable. On the other hand, Morbidelli \& Nesvorny (1999) have shown that due to a process of chaotic diffusion some asteroids in the inner main belt are able to  achieve the Hungaria region. However, the gap of small density in the regions between the 2:3 MMR with Mars and 4:1 MMRs with Jupiter, the $\nu_4$ and $\nu_{16}$ SRs, and with $I>$ 30$^\circ$ (i.e., over the $\nu_5$ SR), seems to define a constrain for the migration paths of asteroids from the inner main belt to this region.

Finally, another interesting result is that we have reporting the first known cases of asteroids  captured in the 3:4 MMR and in the $\nu_4$ SR with Mars, which are (156466) \textit{2002 CG10} and (244666) \textit{2003 JT1}, respectively.

\section{References}

\begin{itemize}

\item Assandri M. C., Gil-Hutton R., 2008, A\&A, 488, 339

\item Bottke W. F., Vokrouhlicky D., Minton D., Nesvorny D., Morbidelli A., Brasser R., Simonson B., Levison H. F., 2012, Nature, 485, 78

\item Brown M. E., Barkume K. M., Ragozzine D., Schaller E. L., 2007, Nature, 446, 294

\item Chirikov B. V., 1979, Phys. Rep., 52, 263–379

\item Cincotta P. M., Simo C., 2000, A\&A, 147, 205

\item Connors M., Greg Stacey R., Wiegert P., Brasser R., 2008, Icarus, 194, 789

\item Correa Otto J. A., Leiva A. M., Giuppone C. A., Beauge C., 2010, MNRAS,
402, 1959

\item Correa-Otto J., Michtchenko T. A., Beauge C., 2013, A\&A, 560, A65

\item Cuk M., 2012, Icarus, 218, 69

\item Cuk M., Nesvorny D., 2018, Icarus, 304, 9

\item Farinella P., Vokrouhlicky D., 1999, Science, 283, 1507

\item Froeschle C., Morbidelli A., 1994, IAUS, 160, 189

\item Galiazzo M. A., Schwarz R., 2014, MNRAS, 445, 3999

\item Gallardo T., Venturini J., Roig F., Gil-Hutton R., 2011, Icarus, 214, 632

\item Gil-Hutton R., Lazzaro D., Benavidez P., 2007, A\&A, 468, 1109

\item Gradie J. C., Chapman C. R., Williams J. G. A., 1979, Families of minor
planets. University of Arizona Press

\item Knezevic Z., Milani A., Farinella P., Froeschle C., Froeschle C., 1991,
Icarus, 93, 316

\item Kozai Y., 1962, AJ, 67, 591

\item Leiva A. M., Correa-Otto J. A., Beauge C., 2013, MNRAS, 436, 3772

\item Lemaitre A., 1994, in Kozai Y., Binzel R. P., Hirayama T., eds, Astronomical Society of the Pacific Conference Series Vol. 63, 75 Years of Hirayama Asteroid Families: The Role of Collisions in the Solar System History. p. 140

\item Lidov M. L., 1961, Iskusst. sputniky Zemly 8, Acad. of Sci., U.S.S.R

\item Lucas M. P., Emery J. P., Pinilla-Alonso N., Lindsay S. S., Lorenzi V., 2017,
Icarus, 291, 268

\item McEachern F. M., Cuk M., Stewart S. T., 2010, Icarus, 210, 644

\item Michel P., Froeschle C., 1997, Icarus, 128, 230

\item Michtchenko T. A., Beauge C., Ferraz-Mello S., 2006, CeMDA, 94, 411

\item Michtchenko T. A., Lazzaro D., Carvano J. M., Ferraz-Mello S., 2010, MNRAS, 401, 2499

\item Migliorini F., Michel P., Morbidelli A., Nesvorný D., Zappala V., 1998,
Science, 281, 2022

\item Milani A., Knezevic Z., Novakovic B., Cellino A., 2010, Icarus, 207, 769

\item Morbidelli A., Henrard J., 1991, CeMDA, 51, 169

\item Morbidelli A., Nesvorny D., 1999, Icarus, 139, 295

\item Murray C. D., Dermott S. F., 1999, Solar system dynamics. Canbridge University Press

\item Ramos X. S., Correa-Otto J. A., Beaugé C., 2015, CeMDA, 123, 453

\item Roig F., Nesvorny D., Gil-Hutton R., Lazaro D., 2008, Icarus, 194, 125

\item Warner B. D., Harris A. W., Vokrouhlicky D., Nesvorny D., Bottke W. F.,
2009, Icarus, 204, 172

\item Williams J., 1989, in Binzel R. P., Gehrels T., Matthews M. S., eds, Asteroids II. pp 1034–1072

\item Williams J. G., 1992, Icarus, 96, 251

\item Williams J. G., Faulkner J., 1981, Icarus, 46, 390

\item Wisdom J., 1980, AJ, 85, 1122

\end{itemize}

\section*{Acknowledgements}

The authors gratefully acknowledge financial support by CONICET through PIP 112-201501-00525. We are grateful to C. Beaug\'e and R. Gil-Hutton by their important suggestions to improve our results. 












\bsp	
\label{lastpage}
\end{document}